\documentclass[sigconf,screen,svgnames,table,hyphens,hidelinks,nonacm]{acmart}

\renewcommand\footnotetextcopyrightpermission[1]{}
\setcopyright{none}
\acmConference[ESEC/FSE 2020]{The 28th ACM Joint European Software
  Engineering Conference and Symposium on the Foundations of Software
  Engineering}{8 - 13 November, 2020}{Sacramento, California, United
  States}

\usepackage{xspace}
\usepackage{booktabs}   %% For formal tables:
\usepackage{subcaption} %% For complex figures with subfigures/subcaptions
\usepackage{xcolor}
\usepackage{textcomp}
\usepackage{pgfpages}
\usepackage{pgfplots}
\usepackage{pgfplotstable}
\usepackage{tikz}
\usepackage{enumerate}
\usepackage{algorithmicx}
\usepackage{algorithm}
\usepackage[noend]{algpseudocode}
\usepackage{multirow}
\usepackage{amsmath}
\usepackage{amsthm}
\usepackage{listings}
\usepackage{colortbl}
\usepackage{url}
\usepackage{titlecaps}
\usepackage{hyperref}
\usepackage[font=bf]{caption}
\usepackage{amssymb}
\usepackage{diagbox}
\usepackage{siunitx}
\usepackage{flushend}
\definecolor{gray}{rgb}{0.5, 0.5, 0.5}
\definecolor{light-gray}{gray}{0.77}
\definecolor{BrickRed}{rgb}{0.8, 0.25, 0.33}
\definecolor{Black}{rgb}{0.0, 0.0, 0.0}
\definecolor{DarkBlue}{rgb}{0.0, 0.0, 0.55}
\definecolor{Crimson}{rgb}{0.86, 0.08, 0.24}
\definecolor{SlateGrey}{rgb}{0.44, 0.5, 0.56}
\definecolor{lightorange}{HTML}{FFB74D}
\definecolor{blue}{rgb}{0.0, 0.0, 1.0}
\definecolor{magenta}{rgb}{0.79, 0.08, 0.48}

\newcommand{\hl}[1]{{\cellcolor[gray]{0.8}} #1}
\newcolumntype{C}{>{\centering}p{1.2cm}}

\makeatletter
\newenvironment{btHighlight}[1][]
{\begingroup\tikzset{bt@Highlight@par/.style={#1}}\begin{lrbox}{\@tempboxa}}
{\end{lrbox}\bt@HL@box[bt@Highlight@par]{\@tempboxa}\endgroup}

\newcommand\btHL[1][]{%
  \begin{btHighlight}[#1]\bgroup\aftergroup\bt@HL@endenv%
}
\def\bt@HL@endenv{%
  \end{btHighlight}%
  \egroup
}
\newcommand{\bt@HL@box}[2][]{%
  \tikz[#1]{%
    \pgfpathrectangle{\pgfpoint{0.3pt}{0pt}}{\pgfpoint{\wd #2}{\ht #2}}%
    \pgfusepath{use as bounding box}%
    \node[anchor=base west,fill=lightorange,outer sep=0pt,inner xsep=0.3pt,inner ysep=0pt,minimum height=\ht\strutbox+0.3pt,#1]{\raisebox{0.3pt}{\strut}\strut\usebox{#2}};
  }%
}
\makeatother

\makeatletter
\newif\if@anonymize

\@anonymizetrue
\@anonymizefalse % Uncomment for camera-ready version

\if@anonymize
  \newcommand\anonymize[1]{Link removed for double-blind review.}
\else
  \newcommand\anonymize[1]{\tiny#1}
\fi
\makeatother
\usetikzlibrary{calc,trees,positioning,arrows,chains,shapes.geometric,%
  decorations.pathreplacing,decorations.pathmorphing,decorations.text,shapes,%
  matrix,shapes.symbols,patterns,shadows,automata}

\def\addlegendimage{\csname pgfplots@addlegendimage\endcsname}
\pgfplotsset{compat=newest}
\newcommand*\LRet[1]{\State \textbf{return} #1}
\newcommand*\Let[2]{\State #1 $\gets$ #2}

\newcommand*\Fcall[1]{\textsc{#1}}

\algrenewcommand\alglinenumber[1]{\tiny\color{Black!70}{#1}}
\algrenewcommand\algorithmicforall[1]{\textbf{foreach} #1}
\algrenewtext{ForAll}{\algorithmicforall}
\algnewcommand\algorithmicswitch{\textbf{switch}}
\algnewcommand\algorithmiccase{\textbf{case}}
\algnewcommand{\LeftComment}[1]{\State \textcolor{SlateGrey}{// #1}}
\algdef{SE}[SWITCH]{Switch}{EndSwitch}[1]{\algorithmicswitch\ #1\ \algorithmicdo}{\algorithmicend\ \algorithmicswitch}%
\algdef{SE}[CASE]{Case}{EndCase}[1]{\algorithmiccase\ #1}{\algorithmicend\ \algorithmiccase}%
\algtext*{EndSwitch}%
\algtext*{EndCase}%
\algdef{SE}[SUBALG]{Indent}{EndIndent}{}{\algorithmicend\ }%
\algtext*{Indent}
\algtext*{EndIndent}
\lstdefinestyle{basic}{%
  alsoletter       = -,%
  morekeywords     = [1]{set-logic,declare-const,declare-fun,assert,let,and,or,not,check-sat,check-sat-using},%
  morekeywords     = [2]{},%
  keywordstyle     = [2]\color{teal}\bfseries,%
  morekeywords     = [3]{},%
  keywordstyle     = [3]\color{BrickRed}\bfseries,%
  keywordstyle     = \bfseries\color{DarkBlue},%
  commentstyle     = \ttfamily\color{Black!70}\lst@ifdisplaystyle\footnotesize\fi,%
  basicstyle       = \ttfamily\lst@ifdisplaystyle\footnotesize\fi,%
  emph             = {Bool,String},%
  emphstyle        = {\color{teal}\bfseries},%
  columns          = [c]fixed,%
  aboveskip        = 0mm,%
  belowskip        = 2mm,%
  keepspaces       = true,%
  mathescape       = true,%
  escapechar       = ¤,%
  tabsize          = 2,%
  numbers          = left,%
  numberstyle      = \tiny\color{Black!70},%
  numbersep        = 4pt,%
  stepnumber       = 1,%
  firstnumber      = 1,%
  showstringspaces = false,%
  captionpos       = b,%
  extendedchars    = true,%
  upquote          = true,%
  abovecaptionskip = 0mm,%
  belowcaptionskip = 0mm,%
  moredelim        = **[is][{\btHL[fill=light-gray]}]{°}{°},%
}

\lstdefinestyle{clang}{%
  language         = C,%
  style            = basic,%
}
\newcommand\code[1]{\lstinline[style=basic]{#1}}

\newcommand\secref[1]{Sect.~\ref{#1}}
\newcommand\figref[1]{Fig.~\ref{#1}}
\newcommand\tabref[1]{Tab.~\ref{#1}}
\newcommand\algoref[1]{Alg.~\ref{#1}}
\newcommand\figsref[1]{Figs.~\ref{#1}}

\newcommand\storm{\textsf{STORM}\xspace}

\newtheoremstyle{mydefinition}%           % Name
  {}%                                     % Space above
  {}%                                     % Space below
  {}%                                     % Body font
  {}%                                     % Indent amount
  {\itshape}%                             % Theorem head font
  {.}%                                    % Punctuation after theorem head
  { }%                                    % Space after theorem head, ' ', or \newline
  {}%

\theoremstyle{mydefinition}

\begin{document}

%% Title information
\title[Detecting Critical Bugs in SMT Solvers Using Blackbox
  Mutational Fuzzing]{Detecting Critical Bugs in SMT Solvers\\Using
  Blackbox Mutational Fuzzing}
                                        %% [Short Title] is optional;
                                        %% when present, will be used in
                                        %% header instead of Full Title.
%% \titlenote{with title note}             %% \titlenote is optional;
%%                                         %% can be repeated if necessary;
%%                                         %% contents suppressed with 'anonymous'
%% \subtitle{Subtitle}                     %% \subtitle is optional
%% \subtitlenote{with subtitle note}       %% \subtitlenote is optional;
%%                                         %% can be repeated if necessary;
%%                                         %% contents suppressed with 'anonymous'

 \author{Muhammad Numair Mansur}
 \affiliation{
   \institution{MPI-SWS \country{Germany}}
 }
 \email{numair@mpi-sws.org}

 \author{Maria Christakis}
 \affiliation{
   \institution{MPI-SWS \country{Germany}}
 }
 \email{maria@mpi-sws.org}

 \author{Valentin W{\"u}stholz}
 \affiliation{
   \institution{ConsenSys \country{Germany}}
 }
 \email{valentin.wustholz@consensys.net}

 \author{Fuyuan Zhang}
 \affiliation{
   \institution{MPI-SWS \country{Germany}}
 }
 \email{fuyuan@mpi-sws.org}

%% Abstract
%% Note: \begin{abstract}...\end{abstract} environment must come
%% before \maketitle command
\begin{abstract}
Formal methods use SMT solvers extensively for deciding formula
satisfiability, for instance, in software verification, systematic
test generation, and program synthesis. However, due to their complex
implementations, solvers may contain critical bugs that lead to
unsound results. Given the wide applicability of solvers in software
reliability, relying on such unsound results may have detrimental
consequences.
In this paper, we present \storm, a novel blackbox mutational fuzzing
technique for detecting critical bugs in SMT solvers. We run our
fuzzer on seven mature solvers and find 29 previously unknown
critical bugs. \storm is already being used in testing new features of
popular solvers before deployment.
\end{abstract}

\maketitle

%% %% 2012 ACM Computing Classification System (CSS) concepts
%% %% Generate at 'http://dl.acm.org/ccs/ccs.cfm'.
%% \begin{CCSXML}
%% <ccs2012>
%% <concept>
%% <concept_id>10002978.10003006.10011634.10011635</concept_id>
%% <concept_desc>Security and privacy~Vulnerability scanners</concept_desc>
%% <concept_significance>500</concept_significance>
%% </concept>
%% <concept>
%% <concept_id>10011007.10011074.10011099.10011102.10011103</concept_id>
%% <concept_desc>Software and its engineering~Software testing and debugging</concept_desc>
%% <concept_significance>500</concept_significance>
%% </concept>
%% </ccs2012>
%% \end{CCSXML}

%% \ccsdesc[500]{Security and privacy~Vulnerability scanners}
%% \ccsdesc[500]{Software and its engineering~Software testing and debugging}
%% %% End of generated code

%% Keywords
%% comma separated list
%% \keywords{keyword1, keyword2, keyword3}  %% \keywords are mandatory in final camera-ready submission

%% \maketitle
%% Note: \maketitle command must come after title commands, author
%% commands, abstract environment, Computing Classification System
%% environment and commands, and keywords command.

%% ----------------------------------------------------------------
\section{Introduction}
\label{sect:intro}
%% ----------------------------------------------------------------

The \emph{Satisfiability Modulo Theories} (SMT)
problem~\cite{BarrettTinelli2018} is the decision problem of
determining whether logical formulas are satisfiable with respect to a
variety of background theories. More specifically, an SMT
\emph{formula} generalizes a Boolean SAT formula by supplementing
Boolean variables with predicates from a set of theories. As an
example, a predicate could express a linear inequality over real
variables, in which case its satisfiability is determined with the
theory of linear real arithmetic. Other theories include bitvectors,
arrays, and integers~\cite{Ganesh2007}, to name a few.

\emph{SMT solvers}, such as CVC4~\cite{BarrettConway2011} and
Z3~\cite{deMouraBjorner2008}, are complex tools for evaluating the
satisfiability of SMT instances. A typical SMT \emph{instance}
contains assertions of SMT formulas and a satisfiability check (see
\figsref{fig:string-bug} and \ref{fig:nuno-bug} for examples). SMT
solvers are extensively used in formal methods, most notably in
software verification (e.g., Boogie~\cite{BarnettChang2005} and
Dafny~\cite{Leino2010}), systematic test case generation (e.g.,
KLEE~\cite{CadarDunbar2008} and Sage~\cite{GodefroidLevin2008}), and
program synthesis (e.g., Alive~\cite{LopesMenendez2015}).
Due to their high degree of complexity, it is all the more likely that
SMT solvers contain correctness issues, and due to their wide
applicability in software reliability, these issues may be detrimental.

\tabref{tab:bug-classes} shows classes of bugs that may occur in SMT
solvers.  We restrict the classification to bugs that manifest
themselves as an incorrect solver result. For bugs in class A, the
solver is \emph{unsound} and returns \texttt{unsat} (i.e.,
unsatisfiable) for instances that are satisfiable. Class B refers to
bugs where the solver returns \texttt{sat} (i.e., satisfiable) for
unsatisfiable instances. A solver is \emph{incomplete} when it returns
\texttt{unknown} for an instance that lies in a decidable theory
fragment. We categorize such bugs in class C. Finally, bugs in class D
indicate crashes where the solver does not return any result.

We call bugs in class A \emph{critical} for two main reasons. First,
such bugs may cause unsoundness in program analyzers that rely on SMT
solvers. As an example, consider a software verifier (e.g.,
Dafny~\cite{Leino2010}) or a test case generator (e.g.,
KLEE~\cite{CadarDunbar2008}) that checks reachability of an error
location by querying an SMT solver. If the solver unsoundly proves
that the error is unreachable (e.g., returns \texttt{unsat} for the
path condition to the error), then the verifier will verify incorrect
code and the testing tool will not generate inputs that exercise the
error.

Second, it is much harder to safeguard against bugs in class A than
bugs in other classes. Specifically, consider that, when an instance
is found to be \texttt{sat}, the solver typically provides a
\emph{model}, that is, an assignment to all free variables in the
instance such that it is satisfiable. Therefore, bugs in class B could
be detected by simply evaluating the instance under the model
generated by the solver (assuming that the model is correct). If this
evaluation returns false, then there is a B bug. Bugs in class C are
detected whenever the solver returns \texttt{unknown} for an instance
that lies in a decidable theory fragment, and bugs in class D are
detected when the solver crashes.

\begin{table}[t]
\vspace{1.5em}
\scalebox{1.00}{
\begin{tabular}{l|cccc}
\toprule
\multicolumn{1}{c|}{\diagbox{\textbf{GT}}{\textbf{SR}}} & \multicolumn{1}{C}{\textbf{\texttt{sat}}} & \multicolumn{1}{C}{\textbf{\texttt{unsat}}} & \multicolumn{1}{C}{\textbf{\texttt{unknown}}} & \multicolumn{1}{C}{\textbf{Crash}} \\
\midrule
\textbf{\texttt{sat}}   & \hl{} & A & C & D \\
\textbf{\texttt{unsat}} & B & \hl{} & C & D \\
\bottomrule
\end{tabular}}
\vspace{0.5em}
\caption{Classes of bugs in SMT solvers. \textbf{GT} stands for ground
  truth and \textbf{SR} for solver result.}
\vspace{-1.5em}
\label{tab:bug-classes}
\end{table}

\paragraph{\textbf{Related work}}
Early work on testing SMT solvers presented
FuzzSMT~\cite{BrummayerBiere2009-FuzzSMT}, a blackbox grammar-based
fuzzer for generating syntactically valid SMT instances (from scratch)
for bitvectors and arrays. Since the satisfiability of the generated
formulas is unknown, the main goal of this fuzzer is to detect crashes
in solver implementations (class D). Critical bugs (class A) may only
be detected with differential testing, when multiple solvers disagree
on the satisfiability of a generated SMT instance.

The above idea was recently extended to enable fuzzing string solvers
by a tool called StringFuzz~\cite{BlotskyMora2018}. Similarly to
FuzzSMT, StringFuzz generates formulas from scratch. In addition, it
can transform existing string instances, but without necessarily
preserving their satisfiability. Consequently, critical bugs in a
single string solver cannot be detected given that the satisfiability
of the formulas is not known a priori---differential testing would
again be needed.

Even more recently, there emerged another technique for testing string
solvers~\cite{BugariuMueller2020}, which synthesizes SMT instances
such that their satisfiability is known by construction. This ground
truth is used to derive test oracles. Violating these oracles
indicates bugs of classes A and B.

Note that finding bugs in classes C and D does not require knowing the
ground truth. As a result, any of the above techniques can in
principle detect such bugs as a by-product.

\paragraph{\textbf{Our approach}}
In this paper, we present a general blackbox fuzzing technique for
detecting critical bugs in any SMT solver. In contrast to existing
work, our technique does not require a grammar to synthesize instances
from scratch. Instead, it takes inspiration from state-of-the-art
mutational fuzzers (e.g., AFL~\cite{AFL}) and generates new SMT
instances by mutating existing ones, called \emph{seeds}. The key
novelty is that our approach generates satisfiable instances from any
given seed. As a result, our fuzzer detects a critical bug whenever an
SMT solver returns \texttt{unsat} for one of our generated
instances. We implement our technique in a tool called \storm, which
has the additional ability to effectively minimize the size of
bug-revealing instances to facilitate debugging.

\paragraph{\textbf{Contributions}}
Our paper makes the following contributions:
\begin{enumerate}
\item We present a novel blackbox mutational fuzzing technique for
  detecting critical bugs in SMT solvers.
\item We implement our technique in a fuzzer that is already being
  used for testing new features of solvers before deployment.
\item We evaluate the effectiveness of our fuzzer on seven mature
  solvers and 43 logics. We found 29 previously unknown critical bugs
  in three solvers (or nine solver variants) and 15 different logics.
\end{enumerate}

\paragraph{\textbf{Outline}}
The next section gives an overview of our
approach. \secref{sect:approach} explains the technical details, and
\secref{sect:implementation} describes our implementation. We present
our experimental evaluation in \secref{sect:experiments}, discuss
related work in \secref{sect:relatedWork}, and conclude in
\secref{sect:conclusion}.

%% ----------------------------------------------------------------
\section{Overview}
\label{sect:overview}
%% ----------------------------------------------------------------

To give an overview of our fuzzing technique for SMT solvers, we first
describe a few interesting examples of \storm in action and then
explain what happens under the hood on a high level.

\paragraph{\textbf{In action}}
One of the critical
bugs\footnote{\anonymize{\url{https://github.com/Z3Prover/z3/issues/2871}}}
found by \storm was in Z3's \texttt{QF\_LIA} logic, which stands for
quantifier-free linear integer arithmetic. We opened a GitHub issue to
report this bug, which resulted in an eight-comment discussion between
two Z3 developers on how to resolve it. The issue was closed but
re-opened a day later with more comments on what still needs to be
fixed. The issue was closed for the last time three days after
that. The fix in Z3 included changing the implementation of Gomory's
cut.

Another critical
bug\footnote{\anonymize{\url{https://github.com/Z3Prover/z3/issues/2994}}}
was detected in Z3's Z3str3 string solver~\cite{BerzishGanesh2017}.
According to a developer of Z3str3, the bug existed for a long time
before \storm found it. During this time, it remained undetected even
though Z3str3 was being tested with fuzzers exclusively targeting
string solvers~\cite{BlotskyMora2018,BugariuMueller2020}. A simplified
version of the SMT instance that revealed the bug is shown on the
right of \figref{fig:string-bug}. (We will discuss it in detail later
in this section.)

A third critical
bug\footnote{\anonymize{\url{https://github.com/Z3Prover/z3/issues/3052}}}
was found in Z3's tactic for applying dominator simplification rules.
The instance that was generated by \storm and revealed the bug spanned
194 lines. The minimization component of \storm reduced this instance
to 17 lines. A simplified version of the instance is shown on the left
of \figref{fig:nuno-bug}. (We discuss it later in this section.)
A developer of the buggy tactic asked us which application generated
this instance, thinking that it was a tool he developed during his
PhD thesis. When we mentioned that it was \storm, he replied
``\emph{What? Your random generator could have done my PhD thesis??
  \&@\#\%, you should have told me sooner :)}''.

\begin{figure*}[t]
\centering
\scalebox{0.9}{
\includegraphics[trim=0cm 14.7cm 15.1cm 0cm, clip]{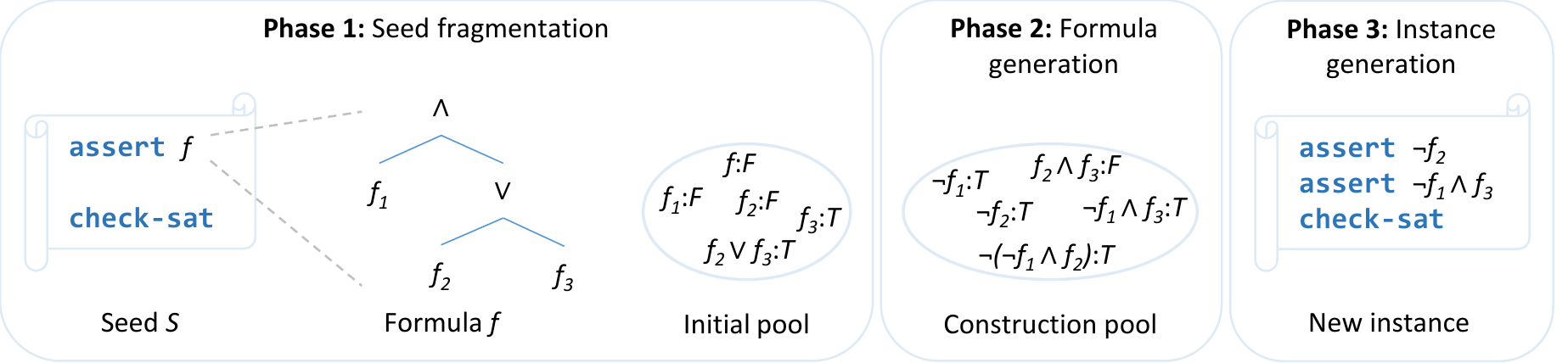}}
%\vspace{-1em}
\caption{Overview of the three \storm phases.}
\label{fig:phases}
\vspace{1em}
\end{figure*}

\begin{figure*}[t]
\begin{minipage}[c]{\columnwidth}
\vspace{-3.275\baselineskip}
\begin{lstlisting}[style=basic]
(declare-const S String)
(assert (str.in.re S (re.++ re.allchar (re.++
  (str.to.re "7;") (re.++ re.allchar
    (str.to.re "aa"))))))
(assert (not (str.in.re S (re.union re.allchar
  (str.to.re "X'jafa")))))
(check-sat)
\end{lstlisting}
\end{minipage}\hfill
\begin{minipage}[c]{\columnwidth}
\begin{lstlisting}[style=basic]
(declare-const S String)
(assert
(let ((a (str.in.re S (re.++ re.allchar (re.++
  (str.to.re "7;") (re.++ re.allchar
    (str.to.re "aa"))))))
(let ((b (not (str.in.re S (re.union re.allchar
  (str.to.re "X'jafa"))))))
(let ((c (and (not b) (not a)))) ¤\label{ln:formula-gen-1}¤
(not c)))))) ¤\label{ln:formula-gen-2}¤
(check-sat)
\end{lstlisting}
\end{minipage}
%\vspace{-1em}
\caption{Original seed instance from SMT-COMP 2019 on the left, and
  simplified instance revealing critical bug in Z3's Z3str3 string
  solver on the right.}
\label{fig:string-bug}
%\vspace{-1em}
\end{figure*}

\begin{figure*}[t]
\begin{subfigure}{\columnwidth}
\begin{lstlisting}[style=basic]
(declare-fun A () Bool)
(declare-fun B () Bool)

(assert (not B))
(assert (not (and (not A) B)))
(assert A)

(check-sat-using dom-simplify)
\end{lstlisting}
\end{subfigure}\hfill
\begin{subfigure}{\columnwidth}
\vspace{-1.645\baselineskip}
\begin{lstlisting}[style=basic]
(declare-fun A () Bool)
(declare-fun B () Bool)

(assert (and (not B) A))

(check-sat-using dom-simplify)
\end{lstlisting}
\end{subfigure}
%\vspace{-1em}
\caption{Simplified instance revealing critical bug in Z3's
  \texttt{dom-simplify} tactic on the left, and logically equivalent
  instance not revealing the bug on the right.}
\label{fig:nuno-bug}
%\vspace{-1em}
\end{figure*}

In \secref{sect:experiments}, we describe in more detail our
experience of using \storm to test both mature solver implementations
as well as new features before their deployment.

\paragraph{\textbf{Under the hood}}
We now give a high-level overview of our fuzzing technique, which
operates in three phases. \figref{fig:phases} depicts each of these
phases.

The first phase, \emph{seed fragmentation}, takes as input a seed SMT
instance $S$. For instance, imagine an instance with multiple
assertions. Each assertion contains a logical formula, such as $f$ in
the figure, potentially composed of Boolean sub-formulas (i.e.,
predicates), such as $f_2 \lor f_3$, $f_1$, $f_2$, and $f_3$ in the
figure. Initially, \storm generates a random assignment of all free
variables in the formulas in $S$. Then, \storm recursively fragments
the formulas in $S$ into all their possible sub-formulas. For example,
$f$ is broken down into $f_1$ and $f_2 \lor f_3$, each of these is in
turn broken down into its Boolean sub-formulas, and so on. The
valuation (i.e., truth value) of each (sub-)formula, $T$ or $F$, is
computed based on the random assignment. All formulas together with
their valuations are inserted in an initial pool as shown in the
figure.

The second phase, \emph{formula generation}, uses the formulas in the
initial pool to build new formulas. The valuation of each new formula
is computed based on the valuations of its constituent initial
formulas. All new formulas with their valuations are inserted in a
construction pool as shown in the figure. For instance, initial
formulas $f_2$ and $f_3$ are used to construct a new formula $f_2
\land f_3$.

The third phase, \emph{instance generation}, uses formulas from both
pools to generate new SMT instances. The reason for having the two
pools is to be able to control the frequency with which initial and
constructed formulas appear in the new instances. Instances generated
during this phase have a different Boolean structure than the
seeds. However, their basic building blocks, that is, the initial
formulas that could not be fragmented further, remain unchanged. This
allows \storm to generate realistic instances as we discuss later in
this section. In addition to being realistic, all new instances are
also satisfiable by construction.

Therefore, a critical bug is detected whenever an SMT solver returns
\texttt{unsat} for a \storm-generated instance. In such a case, \storm
uses \emph{instance minimization} to minimize the size of the instance
revealing the bug.

\paragraph{\textbf{Examples}}
The left of \figref{fig:string-bug} shows a seed instance from the
international SMT competition SMT-COMP 2019~\cite{SMT-COMP}. Starting
from this seed, \storm generated the (simplified) instance on the
right, which revealed the critical bug in Z3str3 described above.
Z3str3 derives length constraints from regular-expression membership
predicates. The bug that \storm exposed here is that such a length
constraint, which is implied by membership in a regular expression,
was not asserted by the string solver.

It is easy to see that the first asserted formula on the left
corresponds to variable \code{a} on the right, while the second
asserted formula on the left corresponds to variable \code{b} on the
right. Therefore, the seed essentially checks for satisfiability of
$\mathtt{a} \land \mathtt{b}$. On the right, \code{c} is equivalent to
$\lnot \mathtt{a} \land \lnot \mathtt{b}$, and the instance checks for
satisfiability of $\lnot \mathtt{c}$, thus, of $\mathtt{a} \lor
\mathtt{b}$. This shows that even small mutations to the Boolean
structure of a formula can be effective in revealing critical issues
in solvers.

This is also evidenced by the example in \figref{fig:nuno-bug}. The
instance on the left reveals the critical bug in Z3's
\texttt{dom-simplify} tactic described earlier. It essentially checks
the satisfiability of $\lnot \mathtt{B} \land \lnot (\lnot \mathtt{A}
\land \mathtt{B}) \land \mathtt{A}$, which is logically equivalent to
$\lnot \mathtt{B} \land \mathtt{A}$. Observe, however, that the
logically equivalent formula, shown on the right of
\figref{fig:nuno-bug}, does not trigger the bug.

Consequently, the benefit of fuzzing the Boolean structure of seed
instances is two-fold. First, it is effective in detecting critical
issues in solvers. Such issues are by definition far more serious and
complex than other types of bugs, such as crashes, since they can, for
instance, result in verifying incorrect safety-critical code. Second,
fuzzing only the Boolean structure of seeds helps generate realistic
SMT instances. This is confirmed by the above comments on the tactic
bug from the Z3 developer who thought that the \storm instance was
generated by his own PhD tool. This was also confirmed by other solver
developers with whom we interacted.

%% ----------------------------------------------------------------
\section{Our approach}
\label{sect:approach}
%% ----------------------------------------------------------------

In this section, we describe our fuzzing technique and how it solves
two key challenges in detecting critical bugs in SMT solvers: (1)~how
to generate non-trivial SMT instances, and (2)~how to determine if a
critical bug is exposed. The latter demonstrates how \storm addresses
the oracle problem~\cite{BarrHarman2015} in the context of soundness
testing for solvers. Finally, we describe how we minimize
bug-revealing instances to reduce their size. This step is crucial for
solver developers as it significantly facilitates debugging.

%% ----------------------------------------------------------------
\subsection{Fuzzing Technique}
\label{subsect:fuzzing-technique}
%% ----------------------------------------------------------------

Given an SMT instance as seed input, our fuzzing approach proceeds in
three main phases: (1)~seed fragmentation, (2)~formula generation, and
(3)~instance generation. Seed fragmentation extracts sub-formulas from
the seed. These will be used as building blocks for generating new
formulas in the second phase. Lastly, instance generation creates new,
satisfiable SMT instances based on the generated formulas, invokes the
SMT solver under test on each of these instances, and uses the solver
result as part of the test oracle to detect critical bugs.

\begin{algorithm}[t]
  \caption{\textbf{Core fuzzing procedure in \storm.}}
  \label{alg:fuzzing}
  % \hspace{-5.65em}\textbf{Input:} Program $\mathit{prog}$, Seeds $S${\btHL[fill=light-gray], Target locations $T$}
  \begin{algorithmic}[1]
    \small
    \Procedure{$\Fcall{PopulateInitialPool}$}{$\mathit{S}, \mathit{D_{max}}$}
      \Let{$\mathit{A}$}{\Fcall{GetAsserts}$(\textit{S})$}
      \Let{$\mathit{M}$}{\Fcall{RandAssignment}$(\textit{A})$} \label{ln:rand-model}
      \Let{$\mathit{P}$}{\Fcall{EmptyPool}$()$}
      \ForAll{$\textit{pred} \in \textit{S}$}
        \If{$\neg\Fcall{ExceedsDepth}(\mathit{pred}, \mathit{D_{max}})$}
          \Let{$\mathit{v}$}{\Fcall{IsTrue}$(\textit{M}, \textit{pred})$}
          \Let{$\mathit{P}$}{\Fcall{Add}$(\textit{P}, \textit{pred}, \textit{v})$}
        \EndIf
      \EndFor
      \LRet{$P$}
    \EndProcedure
    \State
    \Procedure{$\Fcall{Fuzz}$}{$\mathit{S}, \mathit{NC}, \mathit{NM}, \mathit{D_{max}}, \mathit{A_{max}}$}
      \LeftComment{Phase 1: Seed fragmentation}
      \Let{$\mathit{P_{init}}$}{\Fcall{PopulateInitialPool}$(\mathit{S}, \mathit{D_{max}})$} \label{ln:p-init}
      \State
      \LeftComment{Phase 2: Formula generation}
      \Let{$\mathit{P_{constr}}$}{\Fcall{EmptyPool}$()$}
      \While{$\Fcall{Size}(\mathit{P_{constr}}) < \mathit{NC}$}
        \Let{$\mathit{f_1}, \mathit{v_1}$}{\Fcall{RandFormula}$(\textit{P}_{\textit{init}}, \textit{P}_{\textit{constr}})$}
        \Let{$\mathit{op}$}{\Fcall{RandOp}$()$}
        \If{$\mathit{op} = \mathit{AND}$}
          \Let{$\mathit{f_2}, \mathit{v_2}$}{\Fcall{RandFormula}$(\textit{P}_{\textit{init}}, \textit{P}_{\textit{constr}})$} \label{ln:and-start}
          \Let{$\mathit{f}$}{$\mathit{AND}(\mathit{f_1}, \mathit{f_2})$}
          \Let{$\mathit{v}$}{$\mathit{v_1} \wedge \mathit{v_2}$} \label{ln:and-end}
        \Else
          \Let{$\mathit{f}$}{$\mathit{NOT}(\mathit{f_1})$} \label{ln:not-start}
          \Let{$\mathit{v}$}{$\neg\mathit{v_1}$} \label{ln:not-end}
        \EndIf
        \If{$\neg\Fcall{ExceedsDepth}(\mathit{f}, \mathit{D_{max}})$}
          \Let{$\mathit{P_{constr}}$}{\Fcall{Add}$(\textit{P}_{\textit{constr}}, \textit{f}, \textit{v})$}
        \EndIf
      \EndWhile
      \State
      \LeftComment{Phase 3: Instance generation}
      \Let{$\mathit{B}$}{\Fcall{EmptyList}$()$}
      \Let{$\mathit{m}$}{$0$}
      \While{$\mathit{m} < \mathit{NM}$}
        \LeftComment{Number of generated assertions}
        \Let{$\mathit{ac}$}{$(\Fcall{RandInt}()\,\mathit{\%}\,\mathit{A_{max}}) + 1$} \label{ln:ac-def}
        \Let{$\mathit{A}$}{\Fcall{EmptyList}$()$}
        \While{$0 < \mathit{ac}$}
          \Let{$\mathit{f}, \mathit{v}$}{\Fcall{RandFormula}$(\textit{P}_{\textit{init}}, \textit{P}_{\textit{constr}})$}
          \If{$\neg\mathit{v}$}
            \LeftComment{Negation of $\textit{f}\,$ to guarantee assertion satisfiability}
            \Let{$\mathit{f}$}{$\mathit{NOT}(\mathit{f})$}
          \EndIf
          \Let{$\mathit{A}$}{\Fcall{Append}$(\textit{A}, \textit{f})$}
          \Let{$\mathit{ac}$}{$\textit{ac} - 1$}
        \EndWhile

        \LeftComment{Invocation of SMT solver under test}
        \Let{$\mathit{r}$}{\Fcall{CheckSAT}$(\textit{A})$} \label{ln:check-sat}

        \LeftComment{Test oracle}
        \If{$\mathit{r} = \mathit{UNSAT}$}
          \Let{$\mathit{B}$}{\Fcall{Append}$(\textit{B}, \textit{A})$}
        \EndIf

        \Let{$\mathit{m}$}{$\textit{m} + 1$}
      \EndWhile
      \LRet{$B$}
    \EndProcedure
  \end{algorithmic}
  % \hspace{-12.45em}\textbf{Output:} Test suite \textsc{Inputs}$(\mathit{PIDs})$
\end{algorithm}

\algoref{alg:fuzzing} describes these three phases in detail. Function
$\Fcall{Fuzz}$ takes the initial seed $\mathit{S}$ and several
additional parameters that bound the fuzzing process (explained
below). As a first step, the function populates an initial pool
$\mathit{P_{init}}$ of formulas (line~\ref{ln:p-init}) with formula
fragments of the seed $\mathit{S}$.

To this purpose, function $\Fcall{PopulateInitialPool}$ extracts all
assertions in the seed and generates a random assignment $\mathit{M}$,
i.e., an assignment of values to free variables. In our
implementation, we use a separate SMT solver (i.e., different from the
one under test) to generate a model for the assertions (or their
negation if the assertions are unsatisfiable). Next, we iterate over
all predicates (i.e., tree-shaped Boolean sub-formulas as in
\figref{fig:phases}) in the seed. We use assignment $\mathit{M}$ to
evaluate those predicates for which the tree depth does not exceed a
bound $\mathit{D_{max}}$. This valuation $\mathit{v}$ is crucial for
subsequent phases of the fuzzing process, and we add both the formula
$\mathit{pred}$ and $\mathit{v}$ to the initial pool, which is
essentially a map from formulas to valuations. Note that, by
fragmenting the seed, the initial pool already contains a large number
of non-trivial formulas that would be difficult to generate from
scratch (e.g., with a grammar-based fuzzer).

In the second phase, we populate the construction pool
$\mathit{P_{constr}}$ by adding $\mathit{NC}$ new formulas of maximum
depth $\mathit{D_{max}}$. These formulas are generated randomly by
selecting one of two Boolean operators, logical $\mathit{AND}$
(lines~\ref{ln:and-start}--\ref{ln:and-end}) and $\mathit{NOT}$
(lines~\ref{ln:not-start}--\ref{ln:not-end}). Note that this set of
operators is functionally complete, thus allowing us to generate any
Boolean formula. We construct a new formula by conjoining two existing
formulas in the case of $\mathit{AND}$ and negating an existing
formula in the case of $\mathit{NOT}$. Existing formulas are randomly
selected from the pools. Before adding the resulting formula to the
construction pool, we derive its valuation $\mathit{v}$ from the
valuations of its sub-formulas (lines~\ref{ln:and-end}
and~\ref{ln:not-end}).

In essence, the second phase enriches the set of existing formulas by
generating new ones without requiring a complete grammar for all
syntactic constructs. Instead, we use a minimal, but functionally
complete, grammar for \emph{Boolean formulas}. This significantly
simplifies formula generation without sacrificing
expressiveness. Note that a separate pool for newly constructed
formulas allows having control over how many of them are used in the
instances generated in the third phase. On the right of
\figref{fig:string-bug}, this step is responsible for generating the
formulas on lines~\ref{ln:formula-gen-1} and~\ref{ln:formula-gen-2}
that ultimately amount to checking the satisfiability of $\mathtt{a}
\lor \mathtt{b}$.

Once the two pools are populated, we use them to generate
$\mathit{NM}$ SMT instances that we feed to the solver under test. To
assemble a new instance, we create a random number of assertions
($\mathit{ac}$ on line~\ref{ln:ac-def}) by randomly picking formulas
from the pools. If the valuation of a selected formula is true, we
directly assert it, otherwise we assert its negation. This ensures
that all assertions are satisfiable. Of course, the same holds for the
SMT instance consisting of these assertions in addition to a
satisfiability check.
We now leverage this fact when feeding the SMT instance to the solver
under test (line~\ref{ln:check-sat}). The oracle reveals a critical
bug if the solver returns $\mathit{UNSAT}$.

%% ----------------------------------------------------------------
\subsection{Instance Minimization}
\label{subsect:minimization-technique}
%% ----------------------------------------------------------------

In practice, our fuzzing technique often generates bug-revealing
instances that are very large, containing deeply nested formulas and
several assertions. This can considerably complicate debugging for
solver developers.

\begin{algorithm}[t]
  \caption{\textbf{Depth-minimization procedure in \storm.}}
  \label{alg:minimization}
  % \hspace{-5.65em}\textbf{Input:} Program $\mathit{prog}$, Seeds $S${\btHL[fill=light-gray], Target locations $T$}
  \begin{algorithmic}[1]
    \small
    \Procedure{$\Fcall{MinimizeDepth}$}{$\mathit{S}, \mathit{NC}, \mathit{NM}, \mathit{D_{min}}, \mathit{D_{max}}, \mathit{A_{max}}$}
      \If{$\mathit{D_{max}} \leq \mathit{D_{min}}$}
        \LRet{$S$}
      \EndIf

      \Let{$\mathit{D}$}{$(\mathit{D_{min}} + \mathit{D_{max}}) / 2$}
      \Let{$\mathit{B}$}{\Fcall{Fuzz}$(\mathit{S}, \mathit{NC}, \mathit{NM}, \mathit{D}, \mathit{A_{max}})$}
      \If{$0 < \Fcall{Size}(\mathit{B})$}
        \Let{$\mathit{S_{min}}$}{\Fcall{SelectSeedWithSmallestDepth}$(\mathit{B})$}
        \LRet{\Fcall{MinimizeDepth}$(\mathit{S_{min}}, \mathit{NC}, \mathit{NM}, \mathit{D_{min}}, \mathit{D}, \mathit{A_{max}})$} \label{ln:min-depth-call}
      \EndIf
      \LRet{\Fcall{MinimizeDepth}$(\mathit{S}, \mathit{NC}, \mathit{NM}, \mathit{D}+1, \mathit{D_{max}}, \mathit{A_{max}})$}
    \EndProcedure
  \end{algorithmic}
  % \hspace{-12.45em}\textbf{Output:} Test suite \textsc{Inputs}$(\mathit{PIDs})$
\end{algorithm}

% \begin{figure}[t]
% \vspace{0em}
% \center
% % \scalebox{0.6}{
% \footnotesize
% \begin{verbatim}
% func minimizeDepth(S, NUM_CONSTRS, NUM_MUTS,
%   MIN_DEPTH, MAX_DEPTH, MAX_ASSERTS) {
%   if MAX_DEPTH <= MIN_DEPTH {
%     return S
%   }
%   D := (MIN_DEPTH + MAX_DEPTH) / 2
%   bugs := fuzz(S, NUM_CONSTRS, NUM_MUTS,
%     D, MAX_ASSERTS)
%   if 0 < bugs.size() {
%     return minimizeDepth(S, NUM_CONSTRS, NUM_MUTS,
%       MIN_DEPTH, D, MAX_ASSERTS)
%   }
%   return minimizeDepth(S, NUM_CONSTRS, NUM_MUTS,
%     D+1, MAX_DEPTH, MAX_ASSERTS)
% }

% func minimizeAsserts(S, NUM_CONSTRS, NUM_MUTS,
%   MAX_DEPTH, MIN_ASSERTS, MAX_ASSERTS) {
%   ... (analogous to minimizeDepth)
% }
% \end{verbatim}
% % }
% \caption{\storm's minimization algorithm.}
% \label{fig:minimization-algo}
% \vspace{0em}
% \end{figure}

Adapting established minimization techniques based on delta
debugging~\cite{DeltaDebugging} might seem like a natural fit for this
use case. However, the special nature of critical bugs complicates
this task in comparison to other classes of bugs, such as crashes. For
minimizing crashing instances, it is sufficient to minimize the
original instance (e.g., by dropping assertions) \emph{while
  preserving the crash}. In contrast, for instances that exhibit a
critical bug, the behavior that should be preserved is more involved,
that is, \emph{the instance should be minimized such that the buggy
  solver still returns \texttt{unsat} while the ground truth remains
  \texttt{sat}}. This requires either satisfiability-preserving
minimizations or a trusted second solver that can act as a
ground-truth oracle by rejecting minimizations that do not preserve
satisfiability. Unfortunately, the only state-of-the-art delta
debugger for SMT instances, ddSMT~\cite{NiemetzBiere2013}, does not
preserve satisfiability. (Note that ddSMT is the successor of
deltaSMT~\cite{DeltaSMT}, which was used to minimize instances
generated by FuzzSMT~\cite{BrummayerBiere2009-FuzzSMT}.) Moreover, a
second trusted solver is not always available (e.g., for new theories
or solver-specific features and extensions).

To overcome these limitations, we developed a specialized minimization
technique that directly leverages the bounds of our fuzzing procedure
to obtain smaller instances (see \algoref{alg:minimization} for depth
minimization). By repeatedly running the fuzzing procedure on a
\emph{buggy} seed instance, this algorithm attempts to find the
minimum values for $\mathit{D_{max}}$ and $\mathit{A_{max}}$ that
still reveal a critical bug. It uses binary search to first minimize
the number of assertions (analogous to $\Fcall{MinimizeDepth}$ in
\algoref{alg:minimization}) and subsequently the depth of asserted
formulas. Note that the fuzzing procedure may report multiple
bug-revealing instances, and we recursively minimize the smallest with
respect to the bound being minimized
(line~\ref{ln:min-depth-call}). Our evaluation shows that this
technique works more reliably than leveraging ddSMT and a second
solver (see \secref{subsect:results}).

%% ----------------------------------------------------------------
\section{Implementation}
\label{sect:implementation}
%% ----------------------------------------------------------------

\paragraph{\textbf{Seeds}}
\storm uses the Python API in Z3 to manipulate SMT formulas for
generating new instances. It can, therefore, only fuzz instances
within the logics supported by Z3. In practice, this is not an
important restriction since Z3 supports a very large number of
logics. Moreover, \storm requires seeds to be expressed in an
extension of the SMT-LIB v2 input format~\cite{SMT-LIB} supported by
Z3. Note that SMT-LIB is the standard input format used across
solvers.

\paragraph{\textbf{Random assignments}}
Recall that \storm uses Z3 to generate a random model for a given seed
(line~\ref{ln:rand-model} of \algoref{alg:fuzzing}). Note, however,
that bugs in Z3 resulting in a wrong model do not affect our
fuzzer. In fact, given any assignment, our technique just requires correct
valuations for predicates in the initial pool.
In theory, computing these valuations is relatively straightforward
since the assignment provides concrete values for all free variables;
simply substituting variables with values should be sufficient for
quantifier-free predicates. In practice, we use Z3 to compute
predicate valuations and have not encountered any bugs in this solver
component.

\paragraph{\textbf{Random choices}}
Our implementation provides concrete instantiations of functions
$\Fcall{RandOp}$ and $\Fcall{RandFormula}$ from \algoref{alg:fuzzing}
as follows. $\Fcall{RandOp}$ returns $\mathit{AND}$ with probability
50\% and $\mathit{NOT}$ otherwise. Function $\Fcall{RandFormula}$
selects a formula from one of the pools uniformly at random, but with
probability 30\% from the initial pool and from the construction pool
otherwise.

\paragraph{\textbf{Incremental mode}}
Many solvers support a feature called \emph{incremental mode}. It
allows client tools to push and pop constraints when performing a
large number of similar satisfiability queries (e.g., checking
feasibility of paths with a common prefix during symbolic
execution). To efficiently support this mode, solvers typically use
dedicated algorithms that reuse results from previous queries; in
fact, SMT-COMP~\cite{SMT-COMP} features a separate track to evaluate
these algorithms. To test incremental mode, \storm is able to generate
SMT instances that contain push and pop instructions in addition to
regular assertions.

%% ----------------------------------------------------------------
\section{Experimental Evaluation}
\label{sect:experiments}
%% ----------------------------------------------------------------

In this section, we address the following research questions:
\begin{description}
\item[RQ1:] How effective is \storm in detecting new critical
  bugs in SMT solvers?

\item[RQ2:] How effective is \storm in detecting known critical bugs
  in SMT solvers?

\item[RQ3:] How do the assertion and depth bounds of \storm impact its
  effectiveness?

\item[RQ4:] How effective is our instance minimization at reducing the
  size of bug-revealing instances?

\item[RQ5:] To what extent do \storm-generated instances increase code
  coverage of SMT solvers?
\end{description}

%% We make our implementation open
%% source\footnote{\anonymize{\url{https://github.com/Practical-Formal-Methods/storm}}}.
%% To support open science, we commit to providing a publicly available
%% artifact after acceptance of our paper. We will include all data,
%% source code, and documentation necessary for reproducing our
%% experimental results. At the time of submission, we provide both
%%   original and minimized \storm-generated instances for all
%%   bugs in the Supplementary Material.

%% ----------------------------------------------------------------
\subsection{Solver Selection}
\label{subsect:solvers}
%% ----------------------------------------------------------------

We used \storm to test seven popular SMT solvers, which support the
SMT-LIB input format~\cite{SMT-LIB} and regularly participate in the
international SMT competition SMT-COMP~\cite{SMT-COMP}. Specifically,
we selected Boolector~\cite{BrummayerBiere2009-Boolector},
CVC4~\cite{BarrettConway2011}, MathSAT5~\cite{CimattiGriggio2013},
SMTInterpol~\cite{ChristHoenicke2012}, STP~\cite{GaneshDill2007},
Yices2~\cite{Dutertre2014}, and Z3~\cite{deMouraBjorner2008}.

In addition to the above mature implementations, \storm was also used
to test new features of solvers. In particular, the developers of
Yices2 asked us to test the new bitvector theory in the MCSAT
solver~\cite{JovanovicBarrett2013} of Yices2, which is based on the
model-constructing satisfiability
calculus~\cite{deMouraJovanovic2013}. MCSAT is an optional component
of Yices2, which is dedicated to quantifier-free non-linear real
arithmetic. \storm did not find bugs in this new theory of MCSAT, and
the theory was integrated with the main version of Yices2 shortly
after. In our experimental evaluation, it is therefore tested as part
of Yices2.

Moreover, the developers of Z3 asked us to test a new arithmetic
solver (let us refer to it as Z3-AS), which they have been preparing
for the last two years. It comes with better non-linear theories and
has just replaced the legacy arithmetic solvers in Z3. According to
the Z3 developers, \storm could help expedite the integration of this
new feature by finding bugs early, which it did. Since Z3-AS has just
now been integrated with the current version of Z3 and we have only
been testing it independently, we include it separately in our
evaluation.

Due to the success of \storm in detecting intricate critical bugs in
Z3-AS, the Z3 developers described our fuzzer as being
``\emph{extremely useful}'' and have now asked us to test Z3's current
debug branch (let us refer to it as Z3-DBG). Z3-DBG implements a
variety of new solver features in which \storm has already detected
a critical bug (see \secref{subsect:results}).

Finally, the developers of the Z3str3 string
solver~\cite{BerzishGanesh2017} asked us to provide them with
\storm-generated string instances. They became aware of \storm since
it detected several critical issues in Z3str3, which we reported. Note
that Z3str3 is developed by the same group of people as
StringFuzz~\cite{BlotskyMora2018}. We, therefore, suspect that \storm
found bugs in Z3str3 that StringFuzz could not find, especially since
StringFuzz does not target critical bugs.
The \storm-generated instances that we provided (in addition to the
bug-revealing ones that we reported) were used as a regression test
suite during the development of performance enhancements in
Z3str3. According to a developer of Z3str3, our instances helped
reveal critical bugs introduced by these enhancements. Most of these
bugs were due to missing or incorrect axioms in Z3str3.

%% ----------------------------------------------------------------
\subsection{Logic Selection}
\label{subsect:logics}
%% ----------------------------------------------------------------

In our experimental evaluation, for each solver, we test the
intersection of all logics supported by the solver, all logics
supported by the SMT-LIB input format, and all logics supported by Z3.
The latter constraint emerges because our implementation relies on Z3
for generating the mutated SMT instances (see
\secref{sect:implementation}).

\tabref{tab:logics} shows the tested logics for each solver. (The
second column and second to last row of the table should be ignored
for now.) The logic abbreviations are explained in the SMT-LIB
standard~\cite{SMT-LIB}, but generally speaking, the following rules
hold. \texttt{QF} stands for quantifier-free formulas, \texttt{A} for
arrays, \texttt{AX} for arrays with extensionality, \texttt{BV} for
bitvectors, \texttt{FP} for floating-point arithmetic, \texttt{IA} for
integer arithmetic, \texttt{RA} for real arithmetic, \texttt{IRA} for
integer real arithmetic, \texttt{IDL} for integer difference logic,
\texttt{RDL} for rational difference logic, \texttt{L} before
\texttt{IA}, \texttt{RA}, or \texttt{IRA} for the linear fragment of
these arithmetics, \texttt{N} before \texttt{IA}, \texttt{RA}, or
\texttt{IRA} for the non-linear fragment, \texttt{UF} for the
extension that allows free sort and function symbols, \texttt{S} for
strings, and \texttt{DT} for datatypes.

\begin{table*}[t]
\scalebox{1.00}{
\begin{tabular}{lrccccccc}
\toprule
& & \multicolumn{7}{c}{\textbf{SMT Solvers}} \\
\multicolumn{1}{c}{\textbf{Logic}} & \multicolumn{1}{c}{\textbf{Seeds}} & Boolector & CVC4 & MathSAT5 & SMTInterpol & STP & Yices2 & Z3 \\
\midrule
{\footnotesize\tt ALIA}       & 42    & & \checkmark & & \checkmark & & & \checkmark \\
{\footnotesize\tt AUFNIA}     & 3     & & \checkmark & & & & & \checkmark \\
{\footnotesize\tt LRA}        & 2444  & & \checkmark & & \checkmark & & & \checkmark \\
{\footnotesize\tt QF\_ALIA}   & 42    & & \checkmark & & \checkmark & & \checkmark & \checkmark \\
{\footnotesize\tt QF\_AUFNIA} & 3     & & \checkmark & \checkmark & & & & \checkmark \\
{\footnotesize\tt QF\_DT}     & 1602  & & \checkmark & & & & & \checkmark \\
{\footnotesize\tt QF\_LRA}    & 1049  & & \checkmark & & \checkmark & & \checkmark & \checkmark \\
{\footnotesize\tt QF\_RDL}    & 261   & & \checkmark & & & & \checkmark & \checkmark \\
{\footnotesize\tt QF\_UFIDL}  & 444   & & \checkmark & & & & \checkmark & \checkmark \\
{\footnotesize\tt QF\_UFNRA}  & 38    & & \checkmark & \checkmark & & & \checkmark & \checkmark \\
{\footnotesize\tt UFDTLIA}    & 327   & & \checkmark & & & & & \checkmark \\
{\footnotesize\tt AUFDTLIA}   & 728   & & \checkmark & & & & & \checkmark \\
{\footnotesize\tt AUFNIRA}    & 1490  & & \checkmark & & & & & \checkmark \\
{\footnotesize\tt NIA}        & 14    & & \checkmark & & & & & \checkmark \\
{\footnotesize\tt QF\_ANIA}   & 8     & & \checkmark & \checkmark & & &  & \checkmark \\
{\footnotesize\tt QF\_AX}     & 555   & & \checkmark & \checkmark & \checkmark & & \checkmark & \checkmark \\
{\footnotesize\tt QF\_FP}     & 40418 & & \checkmark & \checkmark & & & & \checkmark \\
{\footnotesize\tt QF\_NIA}    & 23901 & & \checkmark & \checkmark & & & \checkmark & \checkmark \\
{\footnotesize\tt QF\_S}      & 24323 & & \checkmark & & & & & \checkmark \\
{\footnotesize\tt QF\_UFLIA}  & 580   & & \checkmark & & \checkmark & & \checkmark & \checkmark \\
{\footnotesize\tt UFLIA}      & 9524  & & \checkmark & & \checkmark & & & \checkmark \\
{\footnotesize\tt AUFLIA}     & 3273  & & \checkmark & & \checkmark  & & & \checkmark \\
{\footnotesize\tt BV}         & 5750  & \checkmark & \checkmark  & & & & & \checkmark \\
{\footnotesize\tt NRA}        & 3813  & & \checkmark &  & & & & \checkmark \\
{\footnotesize\tt QF\_AUFBV}  & 49    & \checkmark & \checkmark & & & & \checkmark & \checkmark \\
{\footnotesize\tt QF\_BV}     & 3872  & \checkmark & \checkmark  & & & \checkmark & \checkmark & \checkmark \\
{\footnotesize\tt QF\_IDL}    & 843   & & \checkmark & & \checkmark & & \checkmark & \checkmark \\
{\footnotesize\tt QF\_NIRA}   & 3     & & \checkmark & \checkmark &  & & \checkmark & \checkmark \\
{\footnotesize\tt QF\_UF}     & 7481  & & \checkmark & & \checkmark & & \checkmark & \checkmark \\
{\footnotesize\tt QF\_UFLRA}  & 936   & & \checkmark & & \checkmark & & \checkmark & \checkmark \\
{\footnotesize\tt UF}         & 7596  & & \checkmark & & \checkmark & & & \checkmark \\
{\footnotesize\tt UFLRA}      & 17    & & \checkmark & & \checkmark & & & \checkmark \\
{\footnotesize\tt AUFLIRA}    & 2268  & & \checkmark & & \checkmark & & & \checkmark \\
{\footnotesize\tt LIA}        & 388   & & \checkmark & & \checkmark & & & \checkmark \\
{\footnotesize\tt QF\_ABV}    & 8310  & \checkmark & \checkmark & & & & \checkmark & \checkmark \\
{\footnotesize\tt QF\_AUFLIA} & 1310  & & \checkmark & & \checkmark & & & \checkmark \\
{\footnotesize\tt QF\_BVFP}   & 17196 & & \checkmark & \checkmark & & & & \checkmark \\
{\footnotesize\tt QF\_LIA}    & 2104  & & \checkmark & & \checkmark & & \checkmark & \checkmark \\
{\footnotesize\tt QF\_NRA}    & 4067  & & \checkmark & \checkmark & & & \checkmark & \checkmark \\
{\footnotesize\tt QF\_UFBV}   & 1238  & \checkmark & \checkmark & & & & \checkmark & \checkmark \\
{\footnotesize\tt QF\_UFNIA}  & 478   & & \checkmark & \checkmark & & & \checkmark & \checkmark \\
{\footnotesize\tt UFDT}       & 4527  & & \checkmark & & & & & \checkmark \\
{\footnotesize\tt UFNIA}      & 4446  & & \checkmark & & & & & \checkmark \\
Unsp.                         & 5825  & -- & -- & -- & -- & -- & -- & -- \\
\midrule
\textbf{Total}                & 193586 & 5 & 43 & 10 & 17 & 1 & 19 & 43 \\
\bottomrule
\end{tabular}}
\vspace{0.5em}
\caption{The tested logics per solver and the number of seed instances
  per logic.}
%\vspace{-1em}
\label{tab:logics}
\end{table*}

%% ----------------------------------------------------------------
\subsection{Benchmark Selection}
\label{subsect:benchmarks}
%% ----------------------------------------------------------------

For our experiments, we used as seeds all non-incremental SMT-LIB
instances in SMT-COMP 2019~\cite{SMT-COMP}. We also used all SMT-LIB
instances in the regression test suites of CVC4, Yices2, and Z3. The
second column of \tabref{tab:logics} shows how many seeds correspond
to each tested logic. The second to last row of the table (``Unsp.'')
refers to instances in which the logic is unspecified---the solver may
use any.

In general, we only tested each solver with logics, and thus
instances, it supports. For seeds without a specified logic, we only
generated mutations of those that each solver could handle.

%% ----------------------------------------------------------------
\subsection{Experimental Setup}
\label{subsect:setup}
%% ----------------------------------------------------------------

For our experiments, we used the following setting for \storm unless
stated otherwise: $D_{max} = 64$, $A_{max} = 64$, $\mathit{NC}$
between 200 and 1500, and $\mathit{NM}$ between 300 and 1000 (see
Alg.~\ref{alg:fuzzing}).  Both $\mathit{NC}$ and $\mathit{NM}$ were
adjusted dynamically within the above ranges based on the size of the
initial pool. The goal was to use larger values for larger initial
pools, and thus, larger seeds.

We performed all experiments on a 32-core \mbox{Intel}
  \textregistered~Xeon \textregistered~E5-2667 v2 CPU~@~3.30GHz
  machine with 256GB of memory, running Debian GNU/Linux 10 (buster).

\paragraph{\textbf{Comparison with state of the art}}
Except for a single tool~\cite{BugariuMueller2020}, all existing
testing tools for SMT solvers do not use oracles to detect critical
bugs. They, therefore, require differential testing of multiple
solvers to identify such bugs.
In RQ2, we evaluate the effectiveness of \storm at detecting existing
critical bugs, including the only publicly reported one found by the
most closely related tool~\cite{BugariuMueller2020}. Recall that this
tool supports only the theory of strings.

%% ----------------------------------------------------------------
\subsection{Experimental Results}
\label{subsect:results}
%% ----------------------------------------------------------------

We now discuss our experimental results for each of the above research
questions.

\paragraph{\textbf{RQ1: New critical bugs}}
\tabref{tab:bugs} shows critical bugs found by \storm in the SMT
solvers we tested. The first column of the table shows the solvers.
We list Z3str3 separately as it is not the default string solver in
Z3. The second column denotes whether bugs were found in the
incremental mode of a solver, which essentially corresponds to a
different solver variant. The third column lists the logics in which
bugs were found, and the last column shows the number of bugs.
\textbf{Overall, \storm found 29 critical bugs in three mature solvers
  (or nine solver variants) and 15 different logics.}

All of these bugs are previously unknown, unique, and confirmed by the
solver developers. Out of the 29 critical bugs, 19 have
already been fixed in the latest solver versions.
Note that the bugs were only detected by \storm-generated instances,
i.e., none were detected by the seeds.
In addition to the bugs in the table, \storm was also able to detect
known bugs as well as other issues (i.e., of classes C and D) as a
by-product, which we do not report here.

% \storm found eight bugs in the \texttt{QF\_S} logic, six of which were
% detected in Z3str3 and two in Z3's default string solver.As mentioned
% earlier, a recent technique for testing string
% solvers~\cite{BugariuMueller2020} synthesizes SMT instances such that
% their satisfiability is known. We checked whether the eight bugs
% found by \storm also existed in the version of Z3 that this technique
% tested (4.7.1). \todo{We found that all of them did exist in this
%   previous version of Z3}, which means that they were missed by the
% synthesis-based technique. In the next research question, we also
% investigate whether \storm can detect bugs reported by the authors of
% this technique.

The feedback we receive from solver developers is very positive, and
we have been discussing it throughout the paper.  As another example,
a Yices2 developer told us that \storm found real bugs and that it is
especially useful to have the ability to test the incremental mode of
solvers. He also mentioned that they used to run
FuzzSMT~\cite{BrummayerBiere2009-FuzzSMT} on all theories, and that
now this fuzzer runs continuously on new theories generating
``\emph{infinite}'' instances. FuzzSMT, however, does not target
critical bugs, and for this reason, they run VoteSMT~\cite{VoteSMT} to
differentially test solvers and detect incorrect Yices2
results. Despite this, \storm detected four new critical bugs in
Yices2.

Another Yices2 developer commented on the severity of two of the bugs
that \storm found. He mentioned that one was in the pre-processing
component and ``\emph{easy to fix (and an obvious mistake in
  retrospect) but it was in a part of Yices that had probably not been
  exercised much}''. ``\emph{The other one was much more tricky to
  trace and fix, it was related to a combination of features and
  optimization in the E-graph, not localized to a single module}''.

\begin{table}[t]
\scalebox{1.00}{
\begin{tabular}{cclc}
\toprule
\multicolumn{1}{c}{\textbf{SMT}} & \multicolumn{1}{c}{\textbf{Incremental}} & \multicolumn{1}{c}{\multirow{2}{*}{\textbf{Logics}}} & \multicolumn{1}{c}{\textbf{Critical}} \\
\multicolumn{1}{c}{\textbf{Solver}} & \multicolumn{1}{c}{\textbf{Mode}} & & \multicolumn{1}{c}{\textbf{Bugs}} \\
\midrule
\multirow{2}{*}{MathSAT5} & \multirow{2}{*}{} & {\footnotesize\tt QF\_FP} & \multirow{2}{*}{2} \\
& & {\footnotesize\tt QF\_BVFP} & \\
\midrule
\multirow{2}{*}{Yices2} & \multirow{2}{*}{} & {\footnotesize\tt QF\_UFIDL} & \multirow{2}{*}{2} \\
& & {\footnotesize\tt QF\_UF} & \\
\midrule
\multirow{2}{*}{Yices2} & \multirow{2}{*}{\checkmark} & {\footnotesize\tt QF\_UFIDL} & \multirow{2}{*}{2} \\
& & {\footnotesize\tt QF\_UFLRA} & \\
\midrule
\multirow{6}{*}{Z3} & \multirow{6}{*}{} & {\footnotesize\tt QF\_UFLIA} & \multirow{6}{*}{8} \\
& & {\footnotesize\tt QF\_BV} & \\
& & {\footnotesize\tt UF} & \\
& & {\footnotesize\tt LIA} & \\
& & {\footnotesize\tt QF\_BVFP} & \\
& & {\footnotesize\tt QF\_LIA} & \\
\midrule
\multirow{2}{*}{Z3} & \multirow{2}{*}{\checkmark} & {\footnotesize\tt QF\_FP} & \multirow{2}{*}{3} \\
& & {\footnotesize\tt QF\_S} & \\
\midrule
Z3str3 & & {\footnotesize\tt QF\_S} & 6 \\
\midrule
\multirow{4}{*}{Z3-AS} & \multirow{4}{*}{} & {\footnotesize\tt AUFNIRA} & \multirow{4}{*}{4} \\
& & {\footnotesize\tt QF\_NIA} & \\
& & {\footnotesize\tt AUFLIRA} & \\
& & {\footnotesize\tt QF\_NRA} & \\
\midrule
Z3-AS & \checkmark & {\footnotesize\tt AUFNIRA} & 1 \\
\midrule
Z3-DBG & & {\footnotesize\tt QF\_NIA} & 1 \\
\bottomrule
\end{tabular}}
\vspace{0.5em}
\caption{Previously unknown, unique, and confirmed critical bugs found
  by \storm in the tested SMT solvers.}
\vspace{-1.5em}
\label{tab:bugs}
\end{table}

\paragraph{\textbf{RQ2: Known critical bugs}}
In this research question, we evaluate the effectiveness of \storm in
reproducing known critical bugs. We, therefore, collected all critical
bugs that were reported for the solvers under test during the
three-month period between Nov 15 and Feb 15, 2020. We focused only on
bugs with a subsequent fix (i.e., closed issues on GitHub). Out of the
seven solvers, we exclude MathSAT5 because it is closed source, and
bugs may only be reported via email. We also exclude Boolector,
SMTInterpol, and STP because no critical bugs were reported for these
solvers during the above time period. For the remaining three solvers,
CVC4, Yices2, and Z3, there were 6, 1, and 14 critical bugs with a
fix, respectively, after excluding all the bugs that we reported.

We ran \storm on the solver version in which each bug was found. Since
developers typically add fixed bugs to their regression tests, we
removed all seeds that revealed any of these bugs (without being
mutated). We collected all generated instances for which each solver
incorrectly returned \texttt{unsat}. To ensure that \storm actually
found the reported bug (and not a different one), we ran all
bug-revealing instances against the first solver version with the
corresponding fix. If the solver now returned \texttt{sat} for at
least one of the instances, we counted the bug as reproduced.

For each of the three solvers, \storm was able to reproduce 1 (CVC4),
1 (Yices2), and 4 (Z3) critical bugs, so 6 out of a total of
21. Therefore, \textbf{if \storm had run on these solver versions, it would
have prevented approximately 1/3 of the critical-bug reports in a
three-month period.} Given that during this period we reported 10
additional bugs detected by \storm in these solvers, it is possible
that our fuzzer would have been able to reproduce more bugs if it had
run longer or if it was being run continuously.

We also tried to find the bugs reported by Bugariu and
M\"uller~\cite{BugariuMueller2020} (regardless of when they were
reported). There was only one GitHub issue opened by these authors
about a critical bug, namely in the Z3str3 string solver. \storm was
able to reproduce this bug.
%
% \todo{Recall, however, that their tool missed the bugs that \storm found (see previous
%   research question).}

%% CVC4: (total: 6) (reproduced: 1)
%% Z3: (total: 21) (reproduced: 11) (7 of them reported by me)
%% Yices: (total: 4) (reproduced: 4) (3 of them reported by me)
%% 15 Nov to 15 Feb
%% + Alexa's single bug in Z3

\paragraph{\textbf{RQ3: Fuzzing bounds}}
To evaluate the effect of the fuzzing bounds of \storm, we only
considered closed bugs. We used all 19 closed bugs reported by us from
RQ1 except for those in Z3-AS (the original commits could not be
retrieved due to a rebase in the branch) for a remaining of 14
bugs. In addition, we used all 7 reproduced bugs (including the one
reported by Bugariu et. al.~\cite{BugariuMueller2020}) from RQ2 for a
total of 21 bugs.

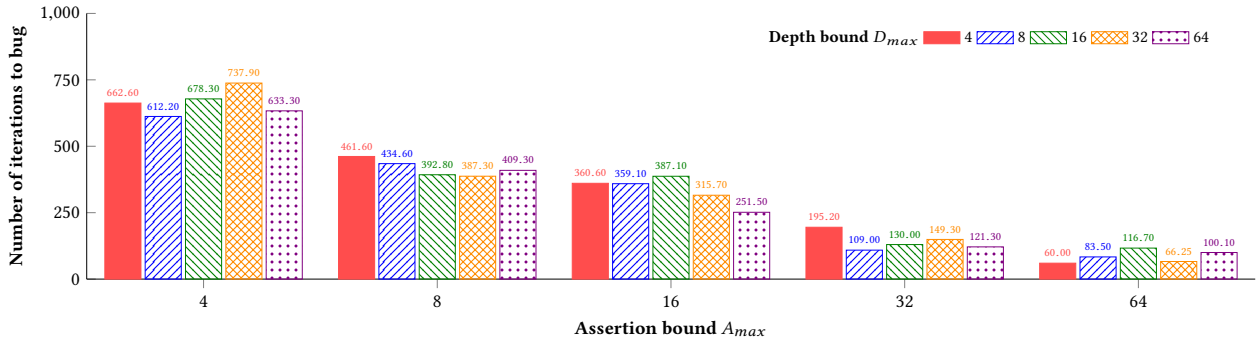
\begin{figure*}[t]
\centering
\scalebox{0.8}{
  \begin{tikzpicture}
    \begin{axis}[
      % ybar stacked,
      ybar,
      enlarge x limits=0.125,
      xticklabels={4, 8, 16, 32, 64},
      xtick=data,
      xlabel={\textbf{Assertion bound $A_{max}$}},
      ylabel={\textbf{Number of iterations to bug}},
      legend cell align=left,
      axis y line*=none,
      axis x line*=bottom,
      width=21cm,
      height=6cm,
      ymin=0,
      ymax=1000,
      ytick distance=250,
      bar width=0.6cm,
      area legend,
      every node near coord/.append style={font=\tiny},
      nodes near coords*={%
        \pgfmathprintnumber[fixed,fixed zerofill, precision=2]\pgfplotspointmeta},
      legend style={legend pos=north east,font=\small, draw=none, legend columns=-1}
      ]

      \addlegendimage{empty legend}
      \pgfplotstableread{results/iterations.dat} \comparison

      \addplot+[ybar, Red!70]
          table[x expr=\coordindex, y=DEP4] from \comparison;

      \addplot+[ybar, Blue, pattern color=Blue, pattern=north east lines]
          table[x expr=\coordindex, y=DEP8] from \comparison;

      \addplot+[ybar, Green, pattern color=Green, pattern=north west lines]
           table[x expr=\coordindex, y=DEP16] from \comparison;

      \addplot+[ybar, DarkOrange, pattern color=DarkOrange, pattern=crosshatch]
          table[x expr=\coordindex, y=DEP32] from \comparison;

      \addplot+[ybar, Purple, pattern color=Purple, pattern=dots]
          table[x expr=\coordindex, y=DEP64] from \comparison;

          \addlegendentry{\hspace{-.6cm}\textbf{Depth bound $D_{max}$}}
          \addlegendentry{4}
          \addlegendentry{8}
          \addlegendentry{16}
          \addlegendentry{32}
          \addlegendentry{64}
    \end{axis}
  \end{tikzpicture}
}
\vspace{-1em}
\caption{Median number of iterations to find bugs with different configurations of
  \storm. Each bar corresponds to a configuration with a certain depth and assertion
  bound.}
\vspace{-1em}
\label{fig:iterations}
\end{figure*}

For each of these bugs we randomly selected a seed file that had
allowed \storm to detect the bug in RQ1 or RQ2. We performed eight
independent runs of \storm (with random seeds different from the ones
used in RQ1 and RQ2 to avoid bias) to evaluate the effect of the different
fuzzing bounds. None of the \storm configurations was able to
reproduce two of the Yices2 bugs from RQ1 with any of the eight random
seeds; we therefore do not include those in the results shown in
Fig.~\ref{fig:iterations}.

\begin{figure*}[t]
\centering
\scalebox{0.8}{
  \begin{tikzpicture}
    \begin{axis}[
      % ybar stacked,
      ybar,
      enlarge x limits=0.125,
      xticklabels={4, 8, 16, 32, 64},
      xtick=data,
      xlabel={\textbf{Assertion bound $A_{max}$}},
      ylabel={\textbf{Time to bug (in seconds)}},
      legend cell align=left,
      axis y line*=none,
      axis x line*=bottom,
      width=21cm,
      height=6cm,
      ymin=0,
      ymax=65,
      ytick distance=10,
      bar width=0.6cm,
      area legend,
      every node near coord/.append style={font=\tiny},
      nodes near coords*={%
        \pgfmathprintnumber[fixed,fixed zerofill, precision=2]\pgfplotspointmeta},
      legend style={legend pos=north east,font=\small, draw=none, legend columns=-1}
      ]

      \addlegendimage{empty legend}
      \pgfplotstableread{results/time.dat} \comparison

      \addplot+[ybar, Red!70]
          table[x expr=\coordindex, y=DEP4] from \comparison;

      \addplot+[ybar, Blue, pattern color=Blue, pattern=north east lines]
          table[x expr=\coordindex, y=DEP8] from \comparison;

      \addplot+[ybar, Green, pattern color=Green, pattern=north west lines]
           table[x expr=\coordindex, y=DEP16] from \comparison;

      \addplot+[ybar, DarkOrange, pattern color=DarkOrange, pattern=crosshatch]
          table[x expr=\coordindex, y=DEP32] from \comparison;

      \addplot+[ybar, Purple, pattern color=Purple, pattern=dots]
          table[x expr=\coordindex, y=DEP64] from \comparison;

          \addlegendentry{\hspace{-.6cm}\textbf{Depth bound $D_{max}$}}
          \addlegendentry{4}
          \addlegendentry{8}
          \addlegendentry{16}
          \addlegendentry{32}
          \addlegendentry{64}
    \end{axis}
  \end{tikzpicture}
}
\vspace{-1em}
\caption{Median time (in seconds) to find bugs with different configurations of
  \storm. Each bar corresponds to a configuration with a certain depth and assertion
  bound.}
\vspace{-0.5em}
\label{fig:time}
\end{figure*}
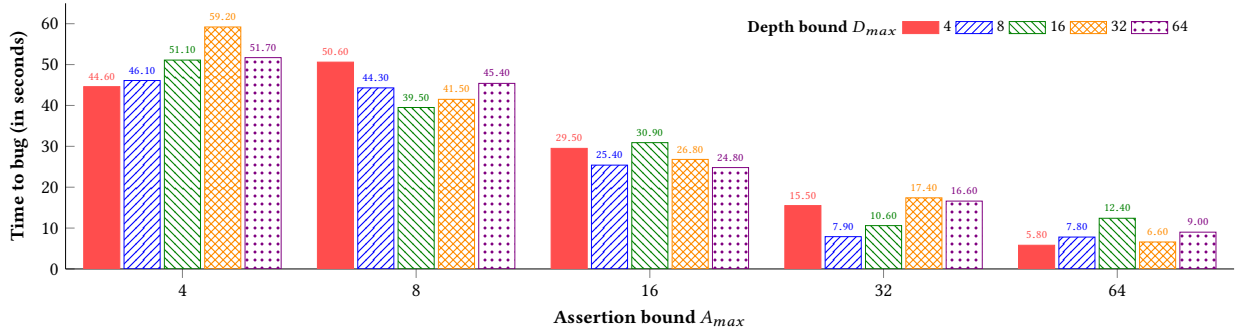

For the assertion and depth bounds $A_{max}$ and $D_{max}$, we used
five different settings: 4, 8, 16, 32, and
64. Fig.~\ref{fig:iterations} shows the median number of iterations
(i.e., generated instances) until the bug was found for different
combinations of these settings. We can observe that \textbf{a large assertion
bound reduces the number of iterations significantly (e.g., up to 11x
for $D_{max} = 4$).} In contrast, the trend for the depth bound is less
clear, which suggests that it has a less significant effect and is
mostly useful for minimizing instances. We can observe very similar
trends when comparing the median time to find the bug (see
Fig.~\ref{fig:time}).

%% 19 own bugs that are closed - 5 of levs (not master) - 2 yices (not reproduced in this experiment) = 12
%% + 6 reproduced bugs
%% + 1 Alexa bug
%% Total: 19

\paragraph{\textbf{RQ4: Instance minimization}}
We now evaluate the effectiveness of our instance minimization. To
this end, we collect all instances revealing the 19 bugs of RQ3 that
are generated by \storm with its default configuration
(\secref{subsect:setup}).

The results of minimizing these instances using binary search (BS) and
delta debugging (ddSMT~\cite{NiemetzBiere2013}) are shown in
\tabref{tab:minimization}. Instance size is measured in terms of the
number of bytes, the number of assertions, and the maximum formula
depth in an assertion. A dash for ddSMT means either that the instance
could not be minimized or that ddSMT does not support a construct in
the instance. As outlined in Sect.~\ref{subsect:minimization-technique}, we had to adapt
ddSMT for this use case by invoking a second solver to reject minimizations that would not
preserve satisfiability; we used the version of the solver that fixed the corresponding
bug for this purpose.

Despite these adaptations, we observed that ddSMT could not minimize
the instances for bugs 2, 3, 4, 5, 13, 14, and 18. We suspect that its
search space of possible minimizations might not contain more complex
transformations that would be required to both preserve satisfiability
\emph{and} the bug. We observed the same outcome when running ddSMT on
instances that were first minimized using binary search.

For bugs 10 and 11, ddSMT does not support \code{str.to.re}, which is
supported by Z3's Z3str3 solver. For the remaining bugs (7, 8, 9, 12,
15, and 16), ddSMT does not support \code{check-sat-using}, which is
supported by Z3. Recall that \storm accepts seed instances expressed
in the extension of the SMT-LIB format that is supported by Z3
(\secref{sect:implementation}), whereas ddSMT only supports the
standard.

Overall, this experiment shows that \textbf{our minimization procedure works
more reliably and is able to significantly reduce buggy instances
(median reduction of 86.1\%).} However, for the cases where both procedures
produced results, the ddSMT-based minimization procedure was able to
produce smaller instances. This is not entirely surprising given that
BS uses the fuzzer, which treats predicates not containing other
predicates (i.e., ground- or leaf-predicates) as atomic building
blocks. For instance, for Bug 17, the instance that was minimized with
BS contains several complex ground-predicates that ddSMT is able to
minimize further. We expect that more involved combinations of the two
approaches could produce even better results.

\paragraph{\textbf{RQ5: Code coverage}}
A Yices2 developer mentioned that they use fuzzer-generated instances
to enrich their regression tests such that they achieve higher
coverage. In this research question, we therefore evaluate whether
\storm is able to increase coverage.

We selected one of the solvers (Z3) and four random logics in which we
found bugs ({\tt QF\_UFLIA, AUFNIRA, UF, LIA}). We then computed the
line and function coverage when running Z3 on all the instances from
SMT-COMP 2019~\cite{SMT-COMP} for these logics (10054 seeds). The
result is shown in the first row of Tab.~\ref{tab:coverage}. At the
same time, we randomly selected 5 instances from each logic and ran
\storm with $NM = 500$ and a single new random seed to generate
exactly 500 new instances for each of the 20 seed
instances. Tab.~\ref{tab:coverage} shows that, as more instances are
generated, coverage increases noticeably (2958 more lines and 1149
more functions after only 500 generated instances). This demonstrates
that \textbf{running \storm on only a small number of seed instances
  is able to result in a noticeable coverage increase over a large
  number of instances from a well known benchmark set.}

\begin{table}[t]
\scalebox{1.00}{
\begin{tabular}{r|rrr|rrr|rrr}
\toprule
\multicolumn{1}{c|}{\textbf{Bug}} & \multicolumn{3}{c|}{\textbf{Unminimized}} & \multicolumn{3}{c|}{\textbf{Minimized}} & \multicolumn{3}{c}{\textbf{Minimized}} \\
\multicolumn{1}{c|}{\textbf{ID}} & \multicolumn{3}{c|}{\textbf{Instances}} & \multicolumn{3}{c|}{\textbf{by BS}} & \multicolumn{3}{c}{\textbf{by ddSMT}} \\
\midrule
 1 & 56832/ & 51/ & 12 & 11687/ & 38/ & 6 & 272/ & 3/ & 3 \\
 2 & 11623/ & 30/ & 2 & 7541/ & 24/ & 2 & --/ & --/ & -- \\
 3 & 15156/ & 30/ & 2 & 5663/ & 24/ & 2 & --/ & --/ & -- \\
 4 & 73656/ & 63/ & 3 & 12983/ & 15/ & 1 & --/ & --/ & -- \\
 5 & 75040/ & 54/ & 10 & 15972/ & 50/ & 3 & --/ & --/ & -- \\
 6 & 54326/ & 27/ & 4 & 8466/ & 6/ & 3 & 507/ & 3/ & 1 \\
 7 & 13160/ & 37/ & 5 & 1253/ & 5/ & 3 & --/ & --/ & -- \\
 8 & 9301/ & 26/ & 2 & 629/ & 3/ & 2 & --/ & --/ & -- \\
 9 & 8209/ & 21/ & 6 & 1138/ & 3/ & 3 & --/ & --/ & -- \\
10 & 401/ & 1/ & 1 & 401/ & 1/ & 1 & --/ & --/ & -- \\
11 & 15322/ & 33/ & 4 & 8342/ & 16/ & 3 & --/ & --/ & -- \\
12 & 13682/ & 40/ & 3 & 282/ & 1/ & 2 & --/ & --/ & -- \\
13 & 8677/ & 57/ & 2 & 660/ & 4/ & 1 & --/ & --/ & -- \\
14 & 12294/ & 32/ & 3 & 1698/ & 6/ & 2 & --/ & --/ & -- \\
15 & 26110/ & 23/ & 4 & 1171/ & 1/ & 2 & --/ & --/ & -- \\
16 & 55423/ & 29/ & 5 & 22660/ & 12/ & 2 & --/ & --/ & -- \\
17 & 96844/ & 55/ & 59 & 2204/ & 4/ & 2 & 260/ & 1/ & 1 \\
18 & 20282/ & 38/ & 4 & 791/ & 4/ & 1 & --/ & --/ & -- \\
19 & 7117/ & 22/ & 4 & 384/ & 1/ & 1 & 172/ & 1/ & 1 \\
\bottomrule
\end{tabular}}
\vspace{0.5em}
\caption{Size of original and minimized bug-revealing
  instances. Instance size is shown in terms of the number of bytes /
  number of assertions / maximum formula depth.}
\vspace{-2em}
\label{tab:minimization}
\end{table}

%% ----------------------------------------------------------------
\subsection{Threats to Validity}
\label{subsect:threats}
%% ----------------------------------------------------------------

We identify the following threats to the validity of our experiments.

\paragraph{\textbf{Selection of seeds}}
\storm requires seed instances as input, and our experimental results
do not necessarily generalize to other
seeds~\cite{SiegmundSiegmund2015}. However, we selected as seeds all
supported instances from SMT-COMP 2019~\cite{SMT-COMP} as well as
regression test suites of solvers. We believe that our selection is
sufficiently broad to mitigate this threat.
%% In addition, we make our
%% tool open source such that it may be run with different seeds.

\paragraph{\textbf{Selection of solvers}}
The bugs found by \storm depend on the solvers and logics that we
tested. However, we selected a wide range of different, mature solvers
and logics to mitigate this threat.

\paragraph{\textbf{Randomness in fuzzing}}
A common threat when evaluating fuzzers is related to the internal
validity~\cite{SiegmundSiegmund2015} of their results. To mitigate
systematic errors that may be introduced due to random choices of our
fuzzer, we used random seeds to ensure deterministic results and
performed experiments for eight different seeds.

%% ----------------------------------------------------------------
\section{Related Work}
\label{sect:relatedWork}
%% ----------------------------------------------------------------

SMT solvers are core components in many program analyzers, and as a
result, their reliability is of crucial importance. Although it is
feasible to verify SAT and SMT
\emph{algorithms}~\cite{FordShankar2002,LescuyerConchon2008,Maric2010},
it is challenging and time consuming to verify even very basic SAT- or
SMT-solver \emph{implementations}. Verifying state-of-the-art,
high-performance solver implementations, such as
CVC4~\cite{BarrettConway2011} and Z3~\cite{deMouraBjorner2008}, is
completely impractical. For these reasons, there is a growing interest
in testing such solvers, alongside related efforts that focus on
testing entire program analyzers.

\paragraph{\textbf{Testing SAT and SMT solvers}}
FuzzSMT~\cite{BrummayerBiere2009-FuzzSMT} focuses on finding crashes
of SMT solvers for bitvector and array instances. It uses
grammar-based blackbox fuzzing to generate crash-inducing instances
and minimizes any such instances with delta
debugging~\cite{DeltaSMT,DeltaDebugging}. Brummayer et
al.~\cite{BrummayerLonsing2010} extend this line of work to SAT and
QBF solvers. In contrast, \storm performs mutational fuzzing, and its
minimization procedure leverages the fuzzer and its bounds regarding
the number of assertions and the formula depth.

StringFuzz~\cite{BlotskyMora2018} targets testing of string
solvers. In addition to randomly generating syntactically valid
instances using a grammar, it is also able to mutate or transform
formulas in existing instances. However, since not all of its
transformations preserve satisfiability, it is not easily possible to
leverage metamorphic testing~\cite{BarrHarman2015} to detect critical
bugs. In contrast to both FuzzSMT and StringFuzz, the satisfiability
of all \storm-generated instances is known.

Recently, Bugariu and M\"uller~\cite{BugariuMueller2020} proposed an
automated testing technique that synthesizes small SMT instances for
the string theory. The true satisfiability of the generated instances
is derived by construction and used as a test oracle. In contrast,
\storm performs mutational fuzzing and supports a wide range of
theories.

Unlike the above approaches that fuzz the input instances of solvers,
Artho et al.~\cite{ArthoBiere2013} and Niemetz et
al.~\cite{NiemetzPreiner2017} develop model-based API testing
frameworks for SAT and SMT solvers. These focus on testing various API
parameters and solver options.

\begin{table}[t]
\scalebox{1.00}{
\begin{tabular}{S[table-format=3.0]cc}
\toprule
\multicolumn{1}{c}{\multirow{1}{*}{\textbf{Generated}}} & \textbf{Line} & \textbf{Function} \\
\textbf{Instances} & \textbf{Coverage} & \textbf{Coverage} \\
\midrule
  0 & 43607 & 21238 \\
100 & 45433 & 22005 \\
200 & 45575 & 22067 \\
300 & 45582 & 22068 \\
400 & 46402 & 22369 \\
500 & 46565 & 22387 \\
\bottomrule
\end{tabular}}
\vspace{0.5em}
\caption{Code coverage increase as more instances are generated by
  \storm.}
\vspace{-2em}
\label{tab:coverage}
\end{table}

\paragraph{\textbf{Testing program analyzers}}
Kapus and Cadar~\cite{KapusCadar2017} combine random program
generation with differential testing~\cite{McKeeman1998} to find bugs
in symbolic-execution engines. Their technique is inspired by existing
compiler-testing techniques (e.g., Csmith~\cite{YangChen2011}) and
used to test KLEE~\cite{CadarDunbar2008}, CREST~\cite{CREST}, and
FuzzBALL~\cite{MartignoniMcCamant2012}.

Cuoq et al.~\cite{CuoqMonate2012} use randomly generated programs to
test the Frama-C static-analysis platform~\cite{FramaC}. Bugariu et
al.~\cite{BugariuWuestholz2018} present a fuzzing technique for
detecting soundness and precision issues in implementations of
abstract domains---the core components of abstract
interpreters~\cite{CousotCousot1977}. They use algebraic properties of
abstract domains as test oracles and find bugs in widely used
domains. Recently, Taneja et al.~\cite{TanejaLiu2020} proposed a
testing technique for identifying soundness and precision issues in
static dataflow analyses by comparing results with a sound and
maximally precise SMT-based analysis; they rely on the SMT solver to
provide correct results.

Zhang et al.~\cite{ZhangSu2019} develop a practical and automated
fuzzing technique to test software model checkers. They focus on
testing control-flow reachability properties of programs. More
specifically, they synthesize valid branch reachability properties
using concrete program executions and then fuse individual properties
of different branches into a single safety property.

Klinger et al.~\cite{KlingerChristakis2019} propose an automated
technique to test the soundness and precision of program analyzers in
general. Their approach is based on differential testing. From seed
programs, they generate program-analysis benchmarks on which they
compare the results of different analyzers.

%% ----------------------------------------------------------------
\section{Conclusion}
\label{sect:conclusion}
%% ----------------------------------------------------------------

In this paper, we have presented a novel fuzzing technique for
detecting critical bugs in SMT solvers---key components of many
state-of-the-art program analyzers. Conceptually, \storm is a blackbox
mutational fuzzer that uses fragments of existing SMT instances to
generate new, realistic instances. Its formula-generation phase takes
inspiration from grammar-based fuzzers; it leverages a minimal, but
functionally complete, grammar for Boolean formulas to generate new
formulas from fragments found in seeds. Finally, it solves the oracle
problem by generating instances that are satisfiable by
construction. \storm found 29 previously unknown critical bugs in
three solvers (or nine solver variants) and 15 different logics.

%% \begin{acks}
%% %
%% We are grateful to the reviewers for their valuable feedback.

%% Maria Christakis's work was supported by DFG grant 389792660 as part
%% of TRR~248 (see \url{https://perspicuous-computing.science}) and a
%% Facebook Faculty Research Award.
%% \end{acks}

%\newpage

%% Bibliography
\bibliographystyle{ACM-Reference-Format} \bibliography{bibliography}

%% Appendix
%\appendix

\end{document}